# Fractional Deviations in Precursor Stoichiometry Dictate the Properties, Performance and Stability of Perovskite Photovoltaic Devices

Paul Fassl[1], Vincent Lami[1], Alexandra Bausch[1], Zhiping Wang[2], Matthew T. Klug[2], Henry J. Snaith[2] and Yana Vaynzof[1*]

[1] P. Fassl, V. Lami, A. Bausch, Prof. Y. Vaynzof, Kirchhoff-Institut für Physik and Centre for Advanced Materials, Ruprecht-Karls-Universität Heidelberg, Im Neuenheimer Feld 227, 69120 Heidelberg, Germany.

[2] **Dr. Z. Wang, Dr. M. T. Klug, Prof. H. J. Snaith,** Clarendon Laboratory, Department of Physics, University of Oxford, Oxford OX1 3PU, UK.

*Email: vaynzof@uni-heidelberg.de

The last five years have witnessed a remarkable progress in the field of lead halide perovskite materials and devices. Examining the existing body of literature reveals staggering inconsistencies in the reported results among different research groups with a particularly wide spread in the photovoltaic performance and stability of devices. In this work we demonstrate that fractional, quite possibly unintentional, deviations in the precursor solution stoichiometry can cause significant changes in the properties of the perovskite layer as well as in the performance and stability of perovskite photovoltaic devices. We show that while the absorbance and morphology of the layers remains largely unaffected, the surface composition and energetics, crystallinity, emission efficiency, energetic disorder and storage stability are all very sensitive to the precise stoichiometry of the precursor solution. Our results elucidate the origin of the irreproducibility and inconsistencies of reported results among different groups as well as the wide spread in device performance even within individual studies. Finally, we propose a simple experimental method to identify the exact stoichiometry of the perovskite layer that researchers can employ to confirm their experiments are performed consistently without unintentional variations in precursor stoichiometry.

**Introduction**

Over the past 5 years lead halide perovskites have been shown to be remarkable materials for application in low cost solar energy devices with efficiencies of up to 22.7 % having been demonstrated to date.[1] Despite tremendous research efforts, one of the key challenges remains the reproducibility and comparability of results between different research groups, with almost all properties of perovskite materials and devices showing large variations when examined by different researchers. While it has been shown that the final quality of perovskite layers, defined for example, by crystallinity, grain size, defect density and morphology is very sensitive to minor variations in the fabrication process,[2,3] accurately following reported experimental procedures does not guarantee similar results.

Many deposition methods have been reported to result in smooth and pinhole-free perovskite films employed in high performing devices. The vast majority of these are based on either a solvent quenching technique, or a two-step process using a mixed perovskite composition of formamidinium (FA), methylammonium (MA), I and Br.[4] For recipes based on the use of antisolvents, the addition of CsI or RbI further improves the performance of devices, often with an additional ~5-10 % excess $PbI_2$ in the precursor solution to achieve best performance.[5,6] The properties of the resulting perovskite film are sensitive to the exact parameters of the antisolvent treatment procedure, such as the choice, and amount of, antisolvent,[7] the timing of the treatment[8] and the local atmosphere in which the washing takes place. In case of the two-step recipe, the final film properties are even more difficult to control since the morphology of the inorganic layer, the reaction time between the conversion solution and precursor layer, as well as the annealing procedure all have a tremendous effect on not only the film microstructure, but also the amount of residual PbI$_2$ in the film.[9,10] For both types of recipes, the exact composition and especially the distribution of the various constituents in the film are hard to regulate and often remain neglected when discussing the results.

Several studies have investigated the effect of stoichiometry on device performance and, in some cases, stability, with conflicting results showing that either excess PbI$_2$[6,11–15] or an excess of anions (I, Br, Cl) and cations (MA, FA, Cs) [8,10,16–18] is beneficial for efficiency and/or stability. The observed discrepancies could stem from the variety of perovskite recipes and the impact of the different extraction layers used in complete cells, as these factors strongly affect the properties of the active layer (e.g. morphology, interfacial charge recombination, stability etc.) and thus may determine which excess is beneficial.[3,19,20] While several recent studies have attempted to link the precursor or surface composition to the microstructure and the electronic and optical properties of perovskite films and devices,[13,19,21–28] the origin for the large variations and difficulties in reproduction of reported results still remains unknown.

In this article, we present a detailed study on how fractional, and quite possibly unintentional, changes in the stoichiometry of the precursor solution have an enormous effect on the properties (surface composition and energetics, photoluminescence, crystallinity, energetic disorder etc.) of perovskite films with seemingly similar, or the same, microstructure and as a result on the device performance and stability. Our study reveals that such fractional changes in the precursor solution could account for the often reported and encountered discrepancies between results of various research groups, and even within the same laboratory. Our work provides vital information on the importance of accurate control and optimization of stoichiometry during fabrication in order to obtain reproducible results compatible with large-scale production and long-term stability, and contributes to further understanding of the sensitive intrinsic properties of this material class.

**Results and Discussion**

**Performing controlled stoichiometric variations**

We employ a recipe based on a lead acetate trihydrate (Pb(Ac)$_2$·3H$_2$O, abbreviated as PbAc$_2$) precursor with the addition of hypophosphorous acid (HPA).[29] It yields very smooth and reproducible high-quality methylammonium lead iodide (MAPbI$_3$) perovskite layers for use in planar perovskite solar cells and has been used in several high impact studies.[30–33] The recipe calls for a molar ratio (denoted as stoichiometry '$y$' throughout the manuscript) of 3:1 of methylammonium iodide (MAI) to PbAc$_2$ ($y = 3.00$). In this recipe, two equivalents of methylammonium acetate, MA(OAc), are removed during the thermal annealing process due to its low thermal stability and the inability of acetate to be incorporated into the perovskite lattice as a result of its small ionic radius.[34,35] Consequently, the PbAc$_2$ recipe results only in phase pure MAPbI$_3$, which is also confirmed by our XRD measurements as described later. However, for a precursor solution amount of 1 ml at 42 wt % (a typical amount for laboratories with small sized substrates), a fractional variation of ≈ 0.5 mg MAI (while weighting 295.5 mg) is sufficient to change the stoichiometry by $\Delta y = 0.005$. Alternatively, a deviation of as little as ≈ 1.5 μL when adding an appropriate amount of MAI/N,N-dimethylformamide (DMF) stock solution to PbAc$_2$ will also result in such a change. Such errors are so small that they commonly occur in laboratory settings. Throughout the study, we vary the stoichiometry from $y = 2.96$ to $y = 3.075$ and perform the variations by adding precisely controlled amounts (in a range of μL) of a MAI/DMF stock solution into the starting perovskite solution (with an accurately weighted stoichiometry $y$). The exact procedure and parameters of solution preparation are explained in greater detail in the Experimental Section and Supplementary Note 1.

**Effect on J-V characteristics of photovoltaic devices**

**Fig. 1a** shows the device structure used in this work. Apart from the fractional variations in precursor stoichiometry all other conditions were kept identical. **Fig. 1b-e** show the photovoltaic parameters of solar cells as a function of precursor stoichiometry (J-V characteristics are shown in **Fig. S1** (ESI†)). Despite only minor changes in stoichiometry, the variations in photovoltaic performance are staggering. The power conversion efficiency (PCE) varies over a range of ~3% reaching a maximum of 15.6 % for $y = 3.055$. Up to $y = 3.055$, both the short circuit current ($J_{SC}$) and the fill factor (FF) remain largely unchanged, however both parameters decrease strongly for $y > 3.055$. The decrease in $J_{SC}$ can also be seen in the external quantum efficiency (EQE) spectra (**Fig. S2a**, ESI†). The open circuit voltage ($V_{OC}$) shows the most interesting trend, with a quasi-linear increase of more than 0.2 V. These trends in the photovoltaic parameters were observed over multiple batches of devices (total number of solar cells: 340) with $2.96 < y < 3.075$ (**Fig. S3**, ESI†) with the best performing devices for $y \approx 3.03$-$3.04$, which is above the typically used $y = 3.00$.

**Effect on surface composition and energetics**

Such remarkable spread in $V_{OC}$ has been reported in literature when comparing the results published by different groups using the same device structure[36] and could have several origins. Optical measurements (**Fig. S2** and **S10,** ESI†) show no change in the active layer absorbance or band gap over the investigated range of stoichiometries, ruling out a $V_{OC}$ increase due to bandgap tuning. A comparison of typical light- and dark J-V curves for various stoichiometries (**Fig. S1**, ESI†) shows an excellent agreement between the increase in $V_{OC}$ and the built-in voltage within the device. An increase in the built-in potential could arise from changes to the electronic structure of the perovskite layer.[37] To probe these changes, we characterized the surface composition and energetics of perovskite/PEDOT:PSS/ITO films using X-ray and Ultraviolet photoemission spectroscopies (XPS and UPS, respectively). **Fig. 2a-c** show the evolution of the Pb4f and I3d doublets and N1s and C1s singlet core levels for various stoichiometries. Apart from the C1s peak at 248.8 eV that is associated with adventitious carbon, all remaining peaks are attributed to $CH_3NH_3PbI_3$. The absence of a peak at ~136.5 eV for Pb4f$_{7/2}$ shows the efficient suppression of metallic lead in our high quality perovskite films for all stoichiometries via an optimized amount of HPA.[29] With increasing stoichiometry, the intensity of the I3d, N1s and C1s peaks increases, resulting in a significant change in the N/Pb and I/Pb atomic ratios (**Fig. 2d**). Olthof et al. have recently shown that MAPbI$_3$ perovskite layers deposited by different methods may result in large deviations in surface composition and ionization potential (IP)[28]. We observe a very similar trend for the IP with changing stoichiometry (**Fig. S4,** ESI†) with values between ~5.7 eV ($y = 3.07$) and ~6.2 eV ($y = 2.97$). Our results indicate that even by using identical fabrication conditions and the same recipe for deposition, the IP of the resulting films can be tuned over a range of 0.5 eV due to fractional variations of the precursor stoichiometry.

We note that the Pb4f$_{7/2}$, I3d$_{5/2}$ and N1s core levels shift to lower binding energies as the precursor stoichiometry increases (**Fig. S4,** ESI†). Comparing the slope of the change in the IP and the shift in the binding energy of the Pb4f$_{7/2}$ peak reveals a direct correlation between these two, which can be explained by the fact that the top of the valence band (VB) is associated to the antibonding interaction between Pb 6s and I 5p atomic orbitals.[38,39] Similarly, the shift in the position of the peak corresponding to MA$^+$ in the UPS spectra w.r.t. vacuum level (VL)[38] directly correlates to the binding energy shifts of N1s and I3d peaks (**Fig. S4,** ESI†). We note that the strong changes in the binding energies described above are not related to changes in the perovskite work function. UPS measurements show that within the experimental error the work function remains unchanged at $5.0 \pm 0.1$ eV (**Fig. 2e**) and is close to the work function measured on PEDOT:PSS/ITO reference samples. Since the work function and the bandgap of the perovskite layers do not change, the VB and conduction band (CB) gradually decrease in energy (w.r.t. VL), suggesting that the surface of the perovskite layer changes from n-type to intrinsic to slightly p-type with increasing stoichiometry. This decrease in the electron affinity of the perovskite layer at the surface allows for an increase in the built-in potential of the device to be formed,[37] as demonstrated schematically in **Fig. S5** (ESI†). We note that the shift in the 'knee' position of the dark J-V characteristics (Fig. S1b, ESI†) directly confirms the increase in the built-in potential of our devices. Since the work function of the layers does not change, the energetic distance between the Fermi level and the CB can thus be can directly correlated to the increased device $V_{OC}$ (**Fig. 2f**).[37] The resulting changes in surface energy levels with varying stoichiometry are illustrated in the inset of **Fig. 2f**, assuming bulk energetics with a stoichiometry of $y = 3.00$. The increase in the built-in potential, and consequently in the $V_{OC}$, originate from the changes at the perovskite/PCBM interface where they are expected to reduce the interfacial recombination. This reduction should drastically increase the electroluminescent (EL) quantum efficiency of the devices.[40] In **Fig. S6** (ESI†) we show the maximum attainable ELQE for various stoichiometries as well as EL spectra measured at a constant current density. As expected, the ELQE increases over three orders of magnitude, indicating that indeed the recombination at the perovskite/PCBM interface is a strongly reduced with increasing stoichiometry.

The remarkable changes in the composition and electronic structure of the surface suggests that the excess MAI from the precursor solution is mainly incorporated into the perovskite lattice in the surface region and possibly in the grain boundaries, while the bulk of the film is composed of a more stoichiometric composition.[16,17,19,21] This is also corroborated by the results of energy-dispersive X-ray (EDX) measurements performed at 6 kV and 10 kV (**Tables S3** and **S4**). Measurements at 6 kV are more representative of film regions closer towards the surface, while those at 10 kV probe the complete bulk.[41] **Fig. S7** (ESI†) shows that the I/Pb atomic ratio, in case of 10 kV, remains largely unchanged with varying precursor stoichiometry, while for 6 kV it increases in a similar fashion to the ratios obtained from XPS.

While it is often claimed that iodine vacancies are the most abundant defects in perovskite films,[42,43] recent reports suggest that under non-equilibrium conditions (taking into account the growth conditions and precursor stoichiometry), iodine and MA interstitials and lead vacancies exhibit the smallest formation energies, especially at the surface.[44–47] We also note that recent experiments by Hawash et al. showed that evaporating thin MAI layers on top of pristine perovskite films similarly results in an increase of the VB position of their samples.[27] The evolution from n-type towards p-type in the case of MAI rich perovskite films has also been shown by Wang et al., albeit for much greater variations in precursor stoichiometry.[48] To summarize, our results indicate that the fractional variations in precursor stoichiometry do not strongly affect the bulk composition of the perovskite layers, but have a tremendous effect on their surface.

**Effect on photovoltaic device stability**

In addition to the large variation in device performance reported by different research groups for seemingly similar device structures and fabrication procedures, large variations in device stability have also been reported.[49] To probe the effect of fractional deviations in stoichiometry on the long-term storage stability of the devices shown in **Fig. 1**, we monitored their performance over 3700 h. The devices were stored without encapsulation in ambient atmosphere (~20-25°, 30-60% RH), and in the dark between the measurements. **Fig. 3** shows the evolution of the photovoltaic parameters and hysteresis index (defined as the ratio of PCE in the reverse and forward scan), and stabilized power output (SPO) measurements after 156 days. The effect on device stability is remarkable. The slightly understoichiometric samples show an incredible storage stability with ~80% of the PCE retained after 3700 h in ambient condition. These devices exhibit nearly no reduction in PCE for the first ~1000 h, with only a 5 % drop of the $J_{SC}$ after 3700 hours. While the FF drops to ~70% of its initial value – probably due to the degradation of the contact layers - the $V_{OC}$ increases in the first ~2200 h.

In stark contrast, even a minute excess of MAI in the precursor solution ($y >= 3.00$) results in significantly reduced stability, with a decrease in the $J_{SC}$ and FF apparent already in the first several hundreds of hours. The SPO measurement in **Fig. 3e** shows that all devices still deliver a stable power output after 156 days, with understoichiometric devices showing a remarkable PCE of nearly 12%, while the previously high efficiency overstoichiometric device with $y = 3.055$ (PCE of 15.6% directly after fabrication) is significantly degraded to a PCE of only 6%. We note that overstoichiometric devices with $y > 3.045$ also develop hysteresis upon storage (**Fig. 3f**). Similar to the device performance variations, these trends in device stability have been observed for multiple batches of devices, with one device with $y = 2.98$ retaining over 80% of its original PCE after 7128 hours of storage in ambient condition which is to our knowledge the highest stability ever reported for a device architecture that includes PEDOT:PSS and Ag electrodes (**Fig. S8** and **S9,** ESI†).

**Effect on microstructure**

Recently it has been shown that device stability, when exposed to $O_2$[50] and $H_2O$,[51] is closely related to the microstructure and average grain size of the perovskite active layer. We characterized the morphology of perovskite/PEDOT:PSS/ITO films with varying stoichiometry by means of scanning electron microscope (SEM) (**Fig. 4a**). For all stoichiometries, we observe pin-hole free perovskite films with grain sizes of ~250 nm with only a slightly reduced average grain size for the highest stoichiometries. For $y < 3.00$ a few white flakes in the areas of the grain boundaries can be observed which have previously been assigned to small amounts of residue $PbI_2$,[14] however these completely vanish for $y > 3.00$. These results show that the fractional stoichiometric variations do not significantly alter the microstructure of the active layer and thus cannot be the cause of the strong variations in device performance and stability.

Many studies employing PEDOT:PSS report rapid performance degradation, often even when stored in inert atmospheres without light illumination.[52–54] Such instability is ascribed to the hygroscopic and acidic nature of PEDOT:PSS. Other studies consistently report that the corrosion of the Ag electrode due to ion migration through $PC_{60}BM$ is one of the main culprits of degradation of the inverted architecture under light illumination,[55–58] or thermal stress[59] occurring even in inert environments and with more stable HTLs. Since in our case all devices had the same structure and were stored in identical conditions, the stark differences in dark storage stability must originate from the fractional variations in precursor stoichiometry.

As we have shown previously, the fractional variations result in strong changes in surface composition, especially in the amount of $MA^+$ and $I^-$ species. The diffusivity and reactivity of these species results in device degradation, which is further accelerated by the presence of moisture, heat or light.[51,55,59–61] Furthermore, based on first principle calculations, MAI terminated surfaces have been suggested to be more prone to water ingress when compared to $PbI_2$ terminated ones.[62,63] So, while slightly overstoichiometric precursor solutions result in favorable surface energetics and an overall higher initial $V_{OC}$ and PCE, they suffer from reduced stability due to the increased amount of $MA^+$ and $I^-$ species at the surface. Interestingly, the remarkable stability of the understoichiometric devices demonstrates that the degradation induced by hygroscopic and acidic PEDOT:PSS and Ag electrodes alone is not the dominant degradation mechanism when stored in ambient atmosphere.

To probe changes in the crystal structure we performed X-ray diffraction (XRD) measurements in $2\theta/\omega$ scan mode, as shown in **Fig. 4b**. All the films show typical tetragonal perovskite diffraction patterns showing two predominant peaks at 14.1° and 28.2° which can be assigned to reflections of the (110) and (220) plane, respectively. For $y < 3.01$ a small amount of $PbI_2$ can be observed at 12.6° supporting the previous assignment of the white flakes in the SEM images to residual $PbI_2$. For $y = 3.07$ a new side peak appears at $2\theta = 11.6°$ which has been previously assigned to a low-dimensional perovskite phase (LDP) in MAI rich films.[64] When we plot the diffraction patterns in logarithmical scale, as shown in **Fig. 4c**, we observe nearly an order of magnitude decrease of the signal intensity for the (110) peak when the stoichiometry changes from $y = 2.97$ to $y = 3.07$. The peak intensity and full width half maximum (FWHM) of the (110) plane plotted over stoichiometry are shown in **Fig. S10** (ESI†). The intensity drops sharply from $y = 2.97$ to $y = 3.00$ and less strongly until $y = 3.07$. This decrease is accompanied by a slight increase in the FWHM, while the lattice parameters remain largely unchanged.

Broadening and shifts in the XRD peak can be induced either by a reduced grain size (Scherrer broadening) or increased microstrain in perovskite films which could influence device efficiency and stability.[30,65,66] To quantify the changes in crystal defects, we performed microstrain analysis using a modified Williamson-Hall method (See Supplementary note 2 for details) with the results shown in **Fig. S11** (ESI†).[30,66] We observe nearly constant microstrain values for the different precursor stoichiometries, suggesting that changes in crystal defects are unlikely to be the origin of the wide variation in performance and stability we describe above. Thus, especially for the range $3.00 < y < 3.05$, we exclude the rather small changes in FWHM and peak intensity as the cause for the strongly reduced device stability we observe. Generally, improved crystallinity has been linked to improved device performance.[10,66] We not only discover that fractional changes in precursor stoichiometry are sufficient to lead to obvious variations in crystallinity, but also that overstoichiometric films despite a lower degree of crystallinity still can deliver higher initial PCE.

**Effect on photoluminescence and energetic disorder**

Emission properties of perovskite films have been extensively studied with conflicting reports regarding the evolution of the photoluminescence quantum efficiency (PLQE) over time and the influence of atmospheric conditions.[31,32,67–70] In **Fig. 5a** we show the evolution of PLQE of MAPbI$_3$ films on glass with stoichiometries fractionally varying around $y = 3.00$. The surrounding atmosphere during the first 20 min is nitrogen, followed by dry air for 10 min. While all films show an enhancement of PLQE in nitrogen, the dynamics of this process is strongly affected by the exact stoichiometry. Slightly overstoichiometric films show the fastest rise time and saturate at a lower PLQE, while the understoichiometric films do not exhibit a stabilized PLQE even after 20 min of illumination. The lower PLQE of overstoichiometric samples is attributed to the increased number of non-radiative decays paths due to a higher amount of iodine at the surface.[45,71,72] In dry air, the variations in PLQE are even more striking:

an increase of the PL occurs for understoichiometric samples (2.985 < $y$ < 3.005), while for $y$ > 3.005 a quenching of the PL can be observed.

Several studies have shown that the PL response to light and various atmospheres on a short timescale (*i.e.* before degradation occurs), is related to trap annihilation either due to exposure to oxygen,[67–70] or due to annihilation of Frenkel defects even in nitrogen atmosphere.[31,32] The exact role of oxygen is still under debate, with recent studies proposing that shallow surface states could be passivated by interaction with oxygen or superoxide species[33,70,72] resulting in an overall enhancement of radiative recombination. Our results unambiguously show that depending on the stoichiometry of the precursor solution and the chemical composition of the surface, either a PL enhancement or quenching can be observed when exposed to oxygen. This reveals that the underlying reasons are very complex and possibly closely related to the exact number of iodine vacancies and interstitials at the surface of the perovskite films.[46,72] We emphasize that even changes as low as $\Delta y = 0.005$ in precursor stoichiometry show a remarkable influence on the PL intensity and dynamics. We believe that monitoring the PL response of perovskite films upon exposure to nitrogen and dry air is a simple method for other research groups to confirm that all their experiments are performed on layers with very similar surface composition. Thus, unintentional variations in the precursor stoichiometry can be identified and eliminated.

The degree of energetic disorder (characterized by Urbach energy $E_u$) of the perovskite films is also strongly influenced by the exact precursor stoichiometry. Photothermal deflection spectroscopy (PDS) measurements (**Fig. S12**, ESI†) reveal that while understoichiometric samples show a similar $E_u$ value of 18-18.5 meV, increasing the precursor stoichiometry beyond $y = 3.00$ results in a continuous increase of $E_u$ up to 21.2 meV for $y = 3.06$ (**Fig. 5b**). This increase could be result of stronger local variations in electrostatic potential caused by different orientations of the polar MA cations[73] or as a result of the increased amount of $MA^+$ and $I^-$ species in the perovskite lattice.[44,45,47] Our results highlight that low $E_u$ is not a guarantee of good photovoltaic performance as previously suggested,[30] since overstoichiometric films with larger values of $E_u$ resulted in higher initial power conversion efficiency when used in devices, however with strongly reduced storage stability.

**Effect on device performance for other perovskite recipes and device architectures**

To explore the universality of the effect of fractional variations in precursor stoichiometry on device performance, we also varied the device architecture (standard vs. inverted) and the perovskite deposition recipe (PbAc$_2$-based vs. PbI$_2$-based). For the PbI$_2$-based recipe, we chose the commonly fabrication route in which the perovskite is derived from of a MAI:PbI$_2$ mixture in DMF:DMSO co-solvent, using a solvent quenching method. Detailed information concerning the methods of solution preparation and device fabrication for each type of device can be found in the Experimental section and Supplementary Note 1. The PV parameters and an illustration of the layer stacks for the four studied device types are presented in **Fig. S13-16** (ESI†). The results show that in all cases, the PCE is strongly affected by the exact precursor stoichiometry, with slight overstoichiometric solutions yielding the best performance for all but one device configuration. The exact effect of the fractional changes in precursor stoichiometry on the photovoltaic parameters depends, unsurprisingly, on the particular device configuration and perovskite recipe and to elucidate the origins of these effects for each recipe/architecture is beyond the scope of this paper. Furthermore, preliminary investigation for devices based on a triple cation perovskite composition (CsFAMA) reveal that also for this recipe small variations in precursor stoichiometry have an effect on device efficiency with a stabilized champion efficiencies of 18.7% for a slight overstoichiometric precursor solution (**Fig. S17**, ESI†). Finally, our preliminary stability measurements performed on PbAc$_2$ devices in the standard architecture show a similar result to the one described above for the inverted structure: the initially best performing overstoichiometric devices with y = 3.03 are strongly degrading after as few as three days, while the slightly understoichiometric devices with y = 2.99 appear to be far more stable (**Fig. S18,** ESI†).

## Conclusions

In conclusion we have revealed the tremendous effect of fractional, and possibly unintentional, variations in the precursor solution on a variety of properties of perovskite films, their photovoltaic performance and stability. Our results suggest that such small variations could account for the often-reported discrepancies in literature and widely varying device performances even when fabricated in the same laboratory, and highlight the critical importance of using the same exact stoichiometry for device fabrication as for film characterization. We suggest that some published results should be carefully examined and rechecked using various perovskite recipes and stoichiometries.

Our work highlights the low tolerance perovskite materials have to small variations in precursor stoichiometry and underlines the need to use high precision and high accuracy instruments (balance, pipettes) with great care when preparing the perovskite solution. We also propose a simple experimental method based on photoluminescence spectroscopy that would allow the identification of systematic errors and should be routinely used as a tool for confirmation that studies are performed on films and devices of the same stoichiometry.

Finally, our results demonstrate that optimization of devices should be performed by carefully fine-tuning the precursor stoichiometry. While reports investigating the effect of precursor ratio changes on the order of 10-20% are common in literature, we show that even a 0.5-1% difference is sufficient to drastically vary the device performance and stability. To perform such studies accurately, we propose that researchers apply our method of using a single perovskite solution, the stoichiometry of which is gradually tuned by adding small amounts of stock solution of only one of the precursors. We hope that our work will help researchers to obtain more reproducible results, as well as help to accomplish a good control over stoichiometries for solution processed devices allowing for further optimization of their efficiency and stability.

**Acknowledgements**

The authors would like to kindly thank Prof. U. Bunz for providing access to device fabrication facilities and Prof. J. Zaumseil for access to SEM/EDX. We are grateful to Prof. Nir Tessler for fruitful discussions. We also thank D. Leibold for help with illustrations and L. Falk for help with device fabrication. P.F. thanks the HGSFP for scholarship. This project has received funding from the European Research Council (ERC) under the European Union's Horizon 2020 research and innovation programme (ERC Grant Agreement n° 714067, ENERGYMAPS).

**Author contributions**

Y.V. and P.F. conceived this study. P.F. designed the experiments, fabricated and measured the devices and performed SEM/EDX measurements. V.L. performed the XPS and UPS measurements and data analysis. A.B. performed the PL and PDS measurements and contributed to device fabrication. Z.W. performed the XRD measurements and data analysis. M. T. K. and Z. W. fabricated and tested the PbI2 based solar cells using the antisolvent technique. Y.V. and H.J.S. guided and supervised the project. P.F and Y.V. jointly wrote the manuscript. All authors discussed the results and reviewed the paper.

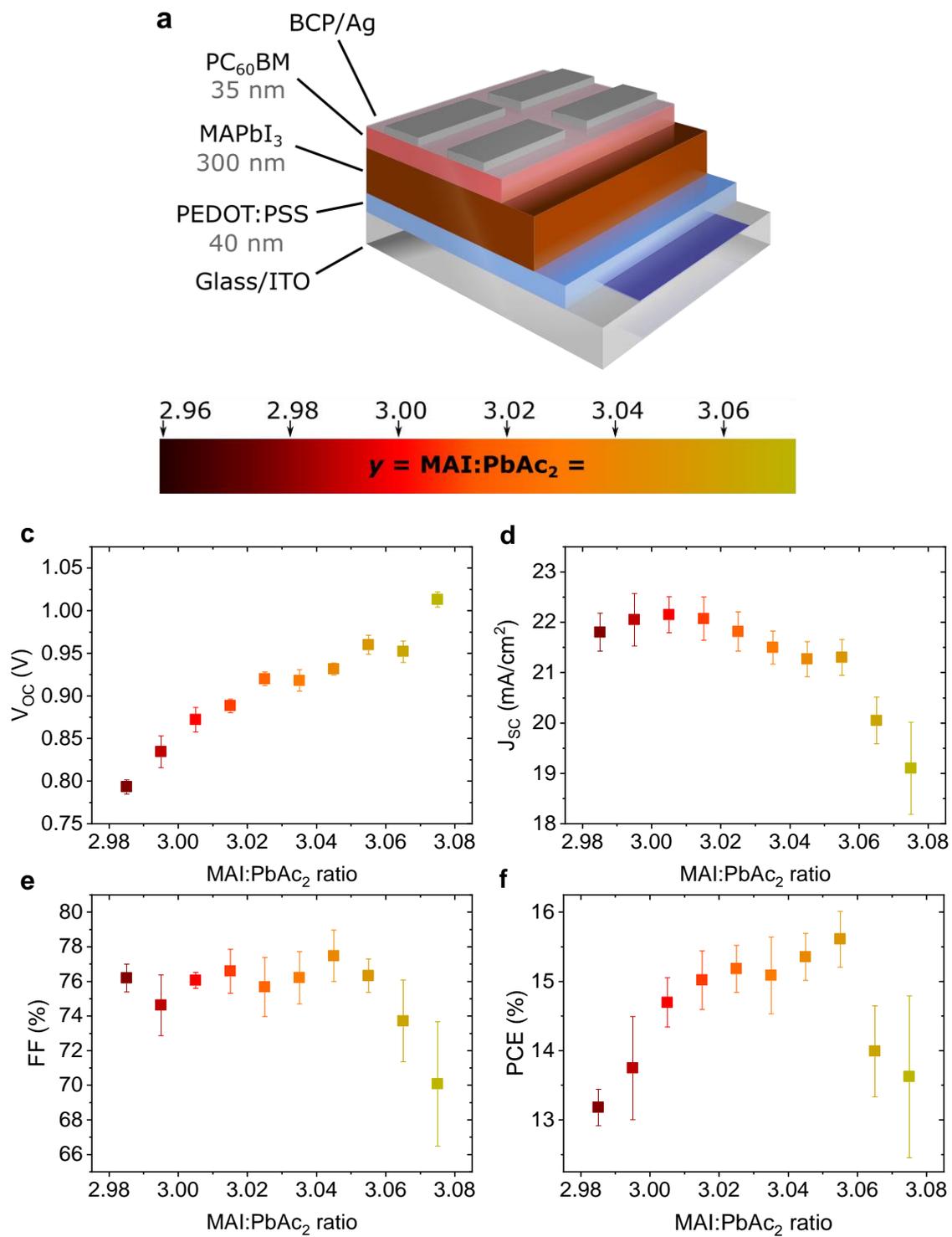

**Figure 1. (a)** Schematic of the photovoltaic device architecture: ITO/PEDOT:PSS/MAPbI$_3$/PCBM/BCP/Ag and the fitting color scale used throughout the manuscript for illustrating stoichiometric changes of the precursor solution, with the stoichiometry $y$ representing the molar ratio of MAI to PbAc$_2$. **(b-e)** Variation of the photovoltaic parameters ($V_{OC}$, $J_{SC}$, FF and PCE) of solar cells with changing stoichiometry ($2.985 < y < 3.075$). Each data represents the averaged value of the reverse and forward scan of ~10-14 cells.

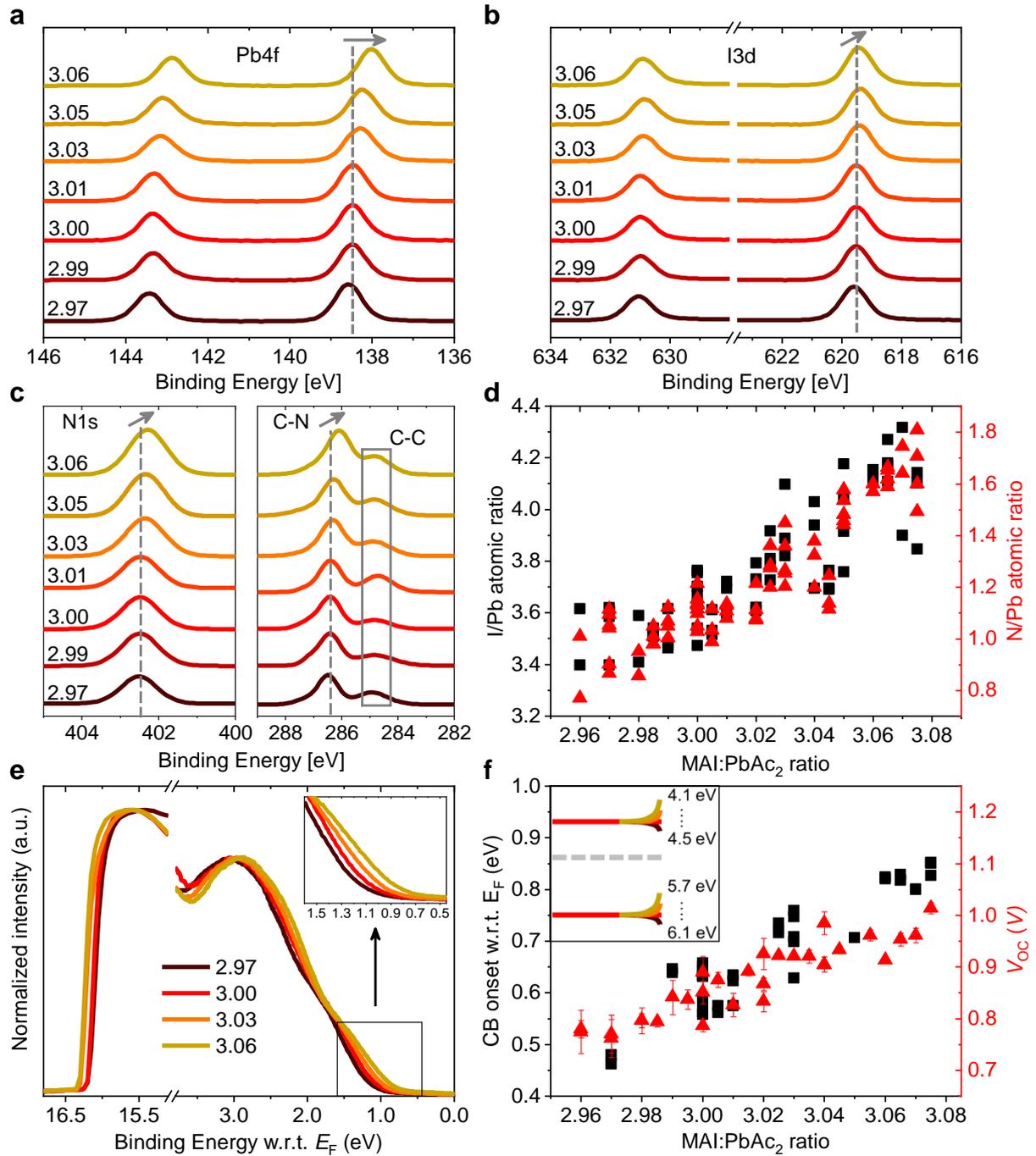

**Figure 2.** Variation of XPS peak positions and intensities of the Pb4f **(a)**, I3d **(b)** and N1s **(c)** peaks with changing stoichiometry. **(d)** I/Pb (left y axis) and N/Pb (right y axis) atomic ratios over stoichiometry determined by XPS. **(e)** Variation of the UPS spectrum for representative stoichiometries w.r.t. the fermi level ($E_F$). The secondary electron onset and the density of states at the valence band (VB) are shown on the left and right, respectively, while the inset is a zoomed in region of the onset of the VB, referencing to the HOMO level. **(f)** Observed change of the conduction band (CB) position w.r.t. $E_F$, calculated from the HOMO level (UPS) and the band gap (UV-vis) and that of the $V_{OC}$ in devices fabricated with various stoichiometries. The inset illustrates the proposed energetic change at the surface of the perovskite films w.r.t. the vacuum level.

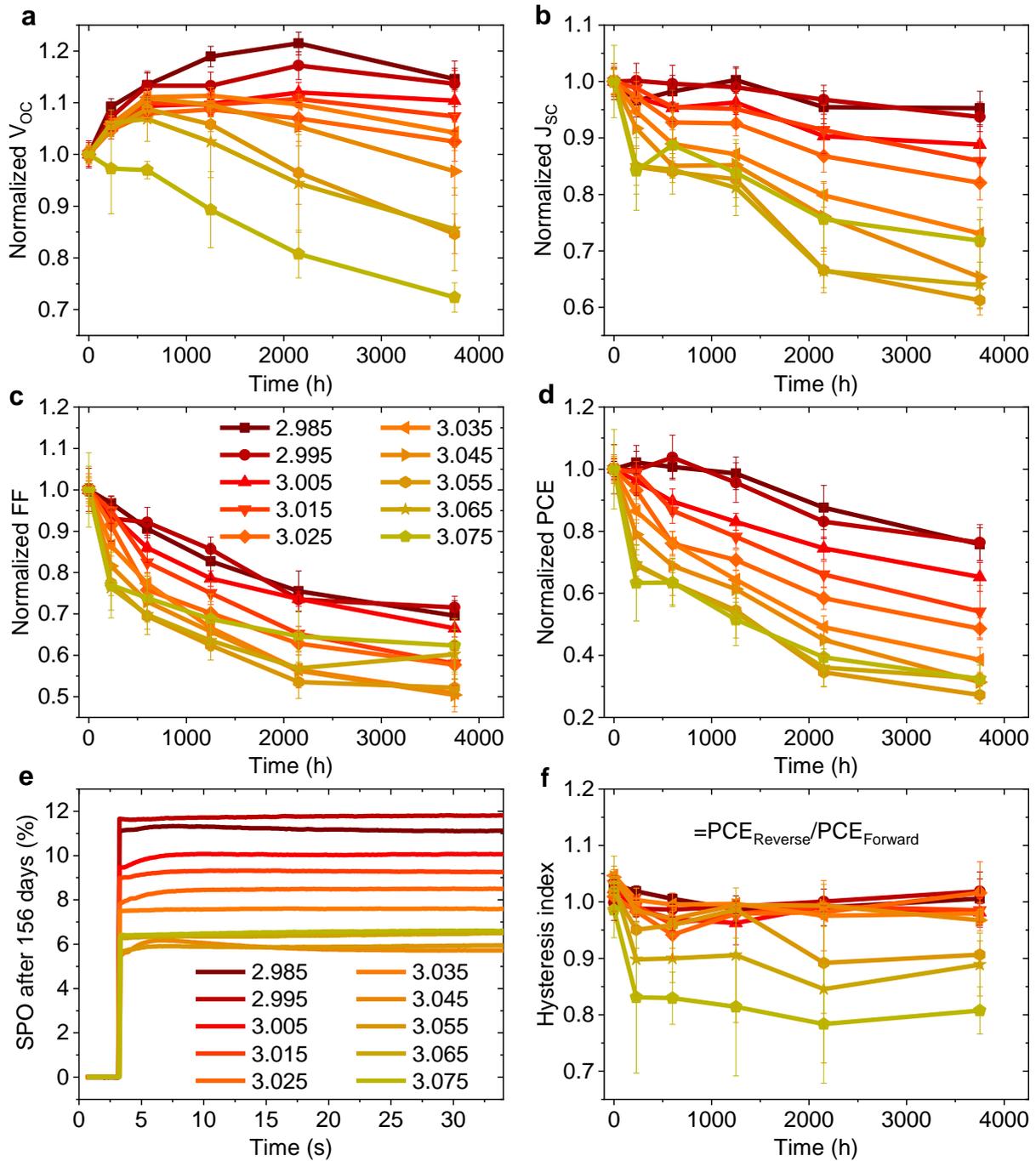

**Figure 3.** (**a-d**) Evolution of photovoltaic parameters in reverse scan of non-encapsulated devices (average of ~8-12 pixels) from the batch shown in **Fig. 1** stored at ~25° and 30-60% RH in the dark. (**e**) Stabilized power output (SPO) of the best pixel for each stoichiometry after 156 days of storage. (**f**) Hysteresis index over time defined by $PCE_{Reverse}/PCE_{Forward}$ of the same devices.

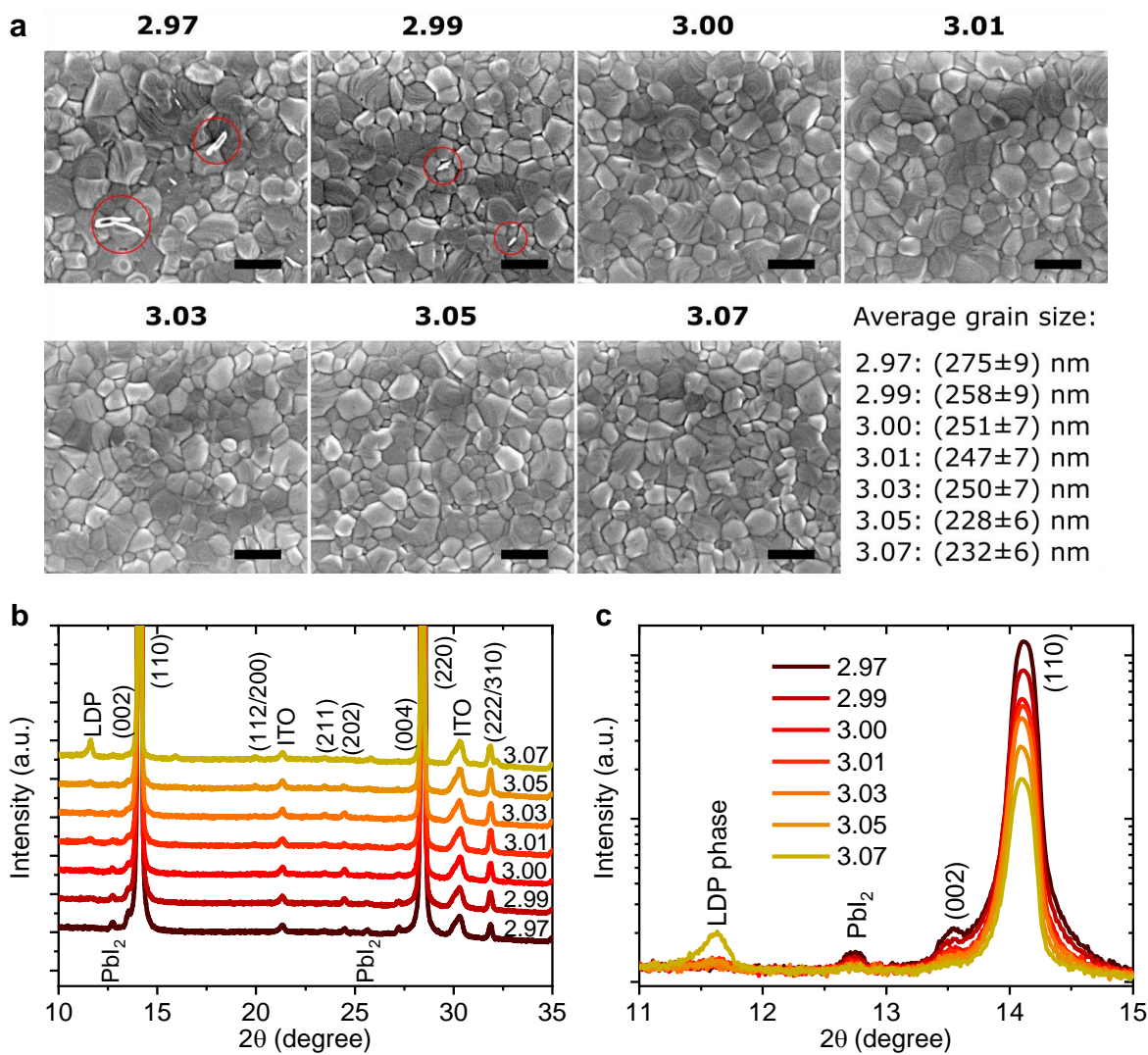

**Figure 4. (a)** Representative SEM images for various stoichiometries as well as averaged grain size of more than 400 individual grains. The white flakes denoted by the red circles are assigned to residual PbI$_2$. The scale bar is 500 nm. **(b)** Linear presentation of XRD spectra for various stoichiometries on ITO/PEDOT:PSS, shifted and magnified for a better visibility of the side peaks. **(c)** Logarithmic presentation of the low angle area of the spectrum presented in **(b)**.

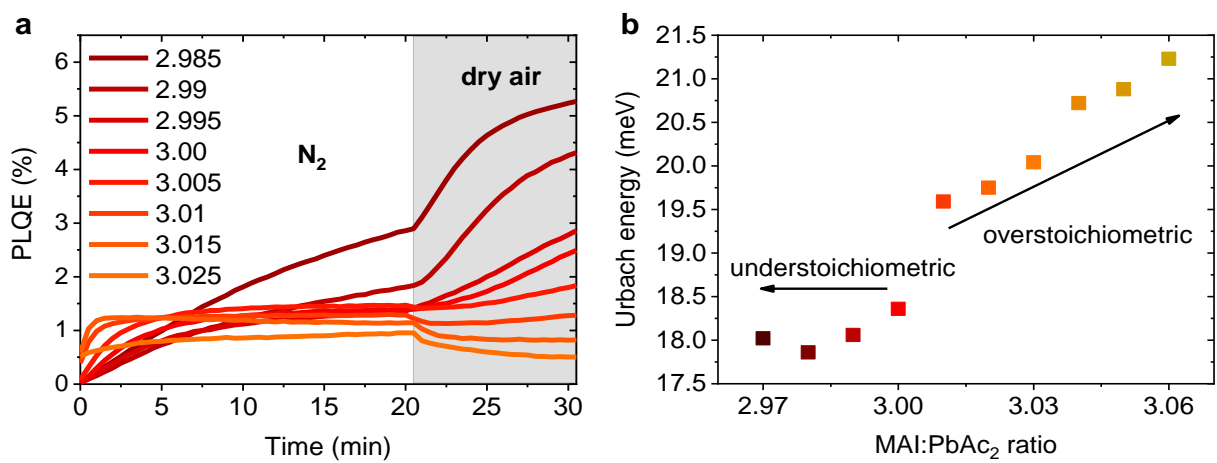

**Figure 5. (a)** Evolution of the photoluminescence quantum efficiency (PLQE) under different atmospheres (0-20 min: N$_2$, 20-30 min: dry air) measured with a 440 nm CW laser (~80 mW/cm$^2$) inside an integrating sphere flushed with the corresponding gas. **(b)** Urbach energy determined by photothermal deflection spectroscopy (PDS).

## Methods

**Materials and Solvents.** CH$_3$NH$_3$I (methylammonium iodide, MAI, M$_w$ = 158.97 g/mol) and HC(NH2)$_2$ (formamidinium iodide, FAI, M$_w$ = 171.91 g/mol) were purchased from GreatCell Solar. CsI (cesium iodide, 99.9 %, M$_w$ = 259.91 g/mol) and PbBr$_2$ (lead bromide, 98+ %, M$_w$ = 367.01 g/mol) were purchased from Alfa Aesar. PbI$_2$ was purchased from TCI. Poly(3,4-ethylenedioxythiophene) poly(styrenesulfonate) (PEDOT:PSS, Al4083) was purchased from Heraeus. Poly-TPD was purchased from 1-Material. 2,2',7,7'-tetrakis[N,N-di(4-methoxyphenyl)amino]-9,9_-spirobifluorene (spiro-OMeTAD) was purchased from Borun Technology. PC$_{60}$BM (>99.5 %) was purchased from Solenne BV. Pb(Ac)$_2$·3(H$_2$O) (lead(II) acetate trihydrate, 99.999 %, PbAc$_2$, M$_w$ = 379.33 g/mol), hypophosphorous acid solution (HPA, 50 wt% in H$_2$O, ρ = 1.206 mg/ml) and all other materials were purchased from Sigma-Aldrich and used as received.

**Perovskite precursor preparation.** For the lead acetate trihydrate (PbAc$_2$) recipe, MAI and PbAc$_2$ in the intended molar ratio (defined as stoichiometry $y$ = MAI:PbAc$_2$) were dissolved in anhydrous N,N-dimethylformamide (DMF) with a concentration of 42 wt%, after which HPA (6.43 µl / 1 ml DMF) was added. The density of this solution was determined by weighting, and the solid concentration of PbAc$_2$ in the solution was calculated, e.g. a density of 1263 mg/ml yields a total (MAI+PbAc$_2$) concentration (41.8 wt% after addition of HPA) of 527.9 mg/ml corresponding to 235.7 mg/ml PbAc$_2$ for a stoichiometry of $y$ = 2.96. Similarly, a MAI/DMF stock solution (29 wt%) with a solid concentration of 332.1 mg/ml MAI was prepared. Appropriate amounts of this stock solution were added to known volumes of perovskite precursor solution to obtain the desired stoichiometries, e.g. 5.95 µl stock solution into 1 ml precursor solution to get from $y$ = 2.96 to $y$ = 2.98 (see Supplementary Note 1 for more details). The perovskite solution for the solvent quenching method (MAPI) recipe was prepared by dissolving 553.2 mg PbI$_2$ and and 181.23 mg MAI in 1 ml anhydrous DMF/DMSO (4:1, v:v) to obtain a 1:0.95 molar ratio solution (PbI$_2$:MAI) with a concentration of 42.96 wt %. A stock solution of MAI in DMF/DMSO (4:1, v:v) with a concentration of 40 wt % was also prepared and added into the perovskite solution in the right amounts to obtain the desired stoichiometries of z = MAI:PbI2 (see Supplementary Note 1 for more details).

The perovskite solution for the triple cation recipe (CsFAMA) recipe was prepared by dissolving 0.948M FAI, 0.192M MAI, 1.02M PbI$_2$, 0.18M PbBr$_2$ and 0.06M CsI in 1 ml anhydrous DMF/dimethyl sulfoxide (DMSO) (4:1, v/v) to obtain stoichiometric Cs$_{0.05}$(FA$_{0.83}$MA$_{0.17}$)$_{0.95}$Pb(I$_{0.9}$Br$_{0.1}$)$_3$ (defined as $z$ = 1) with a concentration of 43.3 wt%. Two stock solutions of Pb(I$_{0.85}$Br$_{0.15}$)$_2$ and FAI$_{0.83}$MAI$_{0.17}$ (both in DMF/DMSO (4:1, v/v) with 31.2 wt% and 18 wt%, respectively) were prepared and added to the perovskite solution in the right amounts to obtain the desired stoichiometries of $z$ = Cs$_x$(FA$_{0.83}$MA$_{0.17}$)$_{100-x}$I:Pb(I$_{0.85}$Br$_{0.15}$)$_2$ (see Supplementary Note 1 for more details). The relative amount of Cesium in the modified precursor solutions decreased by less than 10% after addition of the Cs-free stock solutions for the investigated stoichiometries.

**Device fabrication.** The solar cells in the main text had the architecture of ITO/PEDOT:PSS/CH$_3$NH$_3$PbI$_3$/PCBM/BCP/Ag. Prepatterned indium tin oxide (ITO) coated glass substrates (PsiOTech Ltd., 15 Ohm/sqr) were cleaned sequentially with 2 % hellmanex detergent, deionized water, acetone, and isopropanol, followed by 10 min oxygen plasma treatment. PEDOT:PSS was spin coated at 4000 rpm for 40 s and then annealed at 150 °C for 10 min in ambient air.

The solar cells using Poly-TPD as HTM had the architecture of FTO/Poly-TPD (F4-TCNQ)/CH$_3$NH$_3$PbI$_3$/PCBM/BCP/Ag and were prepared on piranha treated fluorine-doped tin oxide (FTO) coated glass (Pilkington, 15 Ohm/sqr). The HTM solution consisted of 1 mg/ml

Poly-TPD and 0.2 mg/ml F4-TCNQ in Toluene and was spin coated at 2000 rpm for 30 s, followed by annealing at 130° for 10 min in ambient air.

The solar cells using $SnO_2$/PCBM as HTM had the architecture of FTO/$SnO_2$/PCBM/ $CH_3NH_3PbI_3$/spiro-OMeTAD/Ag. Tin (IV) chloride pentahydrate ($SnCl_4 \cdot 5H_2O$; Sigma-Aldrich) was dissolved in anhydrous 2-propanol (0.05M) and stirred for 30 min. The solution was spin-coated at 3000 rpm for 30 s. The substrates were then dried at 100° C for 10 min and annealed at 180° C for 60 min. PCBM was dissolved in anhydrous chlorobenzene (CB; Sigma-Aldrich) at 4 mg/ml and spin-coated on top of the as-prepared $SnO_2$ layer in a nitrogen-filled glove box at 4000 rpm for 40 s and annealed in nitrogen at 70°C for 5 min.

The solar cells using $SnO_2$ nanoparticles ($SnO_2$-NP) as HTM had the architecture of ITO/SnO2-NP/$CH_3NH_3PbI_3$/spiro-OMeTAD/Ag. The $SnO_2$ colloid precursor was obtained from Alfa Aesar (tin(IV) oxide, 15% in $H_2O$ colloidal dispersion). Before use, the particles were diluted by $H_2O$ to 2.67%. The final solution was spin coated onto glass/ITO substrates at 3000 rpm for 30 s and then baked on a hot plate in ambient air at 150° C for 30 min.

The perovskite solution for the $PbAc_2$ recipe was spin coated at 2000 rpm for 60 s in a drybox (RH < 0.5 %). After drying for 5 min at room temperature, the samples were annealed at 100 °C for 5 min.

The perovskite solution for the MAPI recipe was spin coated first at 5000 rpm for 10s with a 6s ramp-up and 6s ramp-down in a nitrogen filled glovebox. 18 s after the spin-coating program began, 300 µL of Anisole was dynamically dispensed onto the wet perovskite film. The film was transferred to a hotplate immediately after spin-coating and annealed for 15 minutes at 100°C.

The perovskite solution for the CsFAMA recipe was spin coated first at 1000 rpm for 15 s and then at 6000 rpm for 30 s in a nitrogen filled glovebox. 200 µL toluene was dripped onto the sample 10 s before the end of the second spinning step. The film was transferred to a hotplate immediately after spin-coating and annealed for 30 min at 100°C.

For the devices in the inverted architecture, PCBM in chlorobenzene (20 mg mL$^{-1}$) was dynamically spin-coated at 2000 rpm for 45 s and annealed at 100 °C for 10 min in a nitrogen filled glovebox. BCP was fully dissolved in isopropanol (0.5 mg mL$^{-1}$) and dynamically spin coated at 4000 rpm for 30 s. For the devices in the standard architecture, spiro-OMeTAD solution was spin-coated on the perovskite layer at 2,500 r.p.m. for 40 s in a dry box (relative humidity <15%) as a hole-transporting layer. To obtain a spiro-OMeTAD solution, we dissolved 85.7mg spiro-OMeTAD (Borun Technology) in 1ml anhydrous chlorobenzene with additives of 28.8 µl tert-butylpyridine (tBP) and 20 µl lithium bis(trifluoromethylsulfonyl)imide (Li-TFSI) salt in acetonitrile (520 mg/ml). Then the devices were kept in a closed container with a relative humidity ~15-20 % overnight to induce oxidisation. To complete the devices, 100 nm silver electrode was deposited via thermal evaporation under high vacuum.

**Current density-voltage (J-V) curves.** The current density-voltage (J-V) measurements were performed under simulated AM 1.5 sunlight at 100 mW cm$^{-2}$ irradiance (Abet Sun 3000 Class AAA solar simulator) with a Keithley 2450 Source Measure Unit. The light intensity was calibrated with a Si reference cell (NIST traceable, VLSI) and corrected by measuring the spectral mismatch between the solar spectrum, the spectral response of the perovskite solar cell and the reference cell. The mismatch factor was calculated to be around 10 %. The cells were scanned from forward bias to short circuit and back at a rate of 0.5 V s$^{-1}$ after holding under illumination at 1.2 V for 2 s. The cells employing PEDOT:PSS as HTL and SnO-NP as ETL were fabricated in Heidelberg with an active area of 0.045 cm$^{-2}$. The cells employing Poly-TPD as HTL and SnO$_2$/PCBM as ETL were prepared in Oxford with an active area of 0.0919 cm$^{-2}$ and a photomask that covers the areas surrounding each pixel from illumination was employed.

**External quantum Efficiency (EQE).** The External Quantum Efficiency (EQE) was measured with the monochromatic light of a halogen lamp from 375 nm to 820 nm, which was calibrated with a NIST-traceable Si diode (Thorlabs).

**Scanning electron microscopy (SEM) and energy dispersive X-ray spectroscopy (EDX).** SEM and EDX analysis were performed using an JSM-7610F FEG-SEM (Jeol). Samples were mounted on standard SEM holders using conductive silver paste to avoid sample charging. For SEM, a working distance of 2 mm and an acceleration voltage of 1.5 kV were used. The size distribution of perovskite grains was estimated using the software ImageJ, analyzing at least 400 grains for each sample. For EDX, a working distance of 15 mm and an acceleration voltage of 6 kV or 10 kV were used. The magnification was chosen low and the integration time was kept short to minimize sample damage.

**Photoemission spectroscopy (XPS/UPS).** The perovskite films investigated for PES measurements were fabricated the same way as the corresponding photovoltaic devices. After the films were spin coated in the dry box, they were stored in a nitrogen glovebox before being transferred into an ultrahigh vacuum (UHV) chamber of the PES system (Thermo Scientific ESCALAB 250Xi) for measurements. The samples were exposed to air only for a short time span of approximately 30 seconds. All measurements were performed in the dark and several spots on each sample were measured in order to ensure enough statistics. UPS measurements were carried out using a double-differentially pumped He discharge lamp (hν = 21.22 eV) with a pass energy of 2 eV and a bias at -10 V. XPS measurements were performed using an XR6 monochromated Al Kα source (hν = 1486.6 eV) and a pass energy of 20 eV.

**UV-Vis and photothermal deflection spectroscopy (PDS).** Optical absorption spectra were measured with a Jasco UV-660 spectrophotometer in the range from 400 to 820 nm. The absorption of the substrate was subtracted as a baseline correction. To correct for scattering effects and for better comparison, the curves were shifted by a constant value to match at 820 nm.
PDS was used to determine Urbach energies of the perovskite films. Perovskite layers for PDS characterization were prepared on spectrosil in an identical way to those on ITO/PEDOT:PSS. The samples were mounted into a sample holder filled with Fluorinert FC-770 (IoLiTec) in a nitrogen filled glovebox. A 150 W Xenon short-arc lamp (Ushio) provides light for a monochromator (Cornerstone 260 Oriel, FWHM 16 nm) to achieve a chopped, tunable, monochromatic pump beam. The heat caused through absorption of the pump light in the perovskite films changes the refractive index of the Fluorinert. This change is detected by deflecting a probe He-Ne-laser (REO) whose displacement in turn is measured by a position-sensitive-detector (Thorlabs, PDP90A). The magnitude of the deflection is determined by a lock-in amplifier (Amatec SR 7230) and directly correlated to the absorption of the film. To estimate the Urbach energies a python leastsquare routine is used to fit an Urbach tail to the measured absorption spectra in the range of the absorption edge.

**X-ray diffraction (XRD).** The X-ray diffraction spectra of the prepared films were measured using a Rigaku SmartLab X-ray diffractometer with CuK$_{\alpha 1}$ (1.54060 Å) and a HyPix-3000 2D hybrid pixel array detector.

**References**
1.  Research Cell Efficiency Records: